\documentclass[prd,twocolumn,showpacs,preprintnumbers,superscriptaddress,floatfix,nofootinbib]{revtex4-1}
\usepackage{amsfonts}
\usepackage{graphicx,,booktabs}
\usepackage{amssymb}
\usepackage{amsmath,txfonts,ulem}
\usepackage{graphicx}
\usepackage{dcolumn}
\usepackage{color}
\usepackage{bm}
\usepackage{ulem}
\usepackage[colorlinks,
            citecolor=blue,
            anchorcolor=green,
            menucolor=orange,
            linkcolor=red,
            filecolor=red,
            runcolor=pink,
            urlcolor=blue,
            frenchlinks=red]{hyperref}

\allowdisplaybreaks[4]

\begin{document}

\title{\boldmath Discovery potential for the LHCb fully-charm tetraquark $X(6900)$
state via $\bar{p}p$ annihilation reaction}
\author{Xiao-Yun Wang}
\email{xywang@lut.edu.cn}
\affiliation{Department of physics, Lanzhou University of Technology,
Lanzhou 730050, China}

\author{Qing-Yong Lin}
\email{qylin@jmu.edu.cn}
\affiliation{Department of Physics, Jimei University,
Xiamen 361021, China}

\author{Hao Xu}
\email{xuh2018@nwpu.edu.cn}
\affiliation{Department of Applied Physics, School of Science, Northwestern Polytechnical University,
Xi'an 710129, China}

\author{Ya-Ping Xie}
\email{xieyaping@impcas.ac.cn}
\affiliation{Institute of Modern Physics, Chinese Academy of Sciences, Lanzhou 730000, China}

\author{Yin Huang}
\email{huangy2019@swjtu.edu.cn}
\affiliation{School of Physical Science and Technology, Southwest Jiaotong University, Chendu 610031, China}

\author{Xurong Chen}
\email{xchen@impcas.ac.cn}
\affiliation{Institute of Modern Physics, Chinese Academy of Sciences, Lanzhou 730000, China}
\affiliation{University of Chinese Academy of Sciences, Beijing 100049, China}
\affiliation{Guangdong Provincial Key Laboratory of Nuclear Science, Institute of Quantum Matter, South China Normal University, Guangzhou 510006, China}

\begin{abstract}
Inspired by the observation of the fully-charm tetraquark $X(6900)$ state at
LHCb, the production of $X(6900)$ in $\bar{p}p\rightarrow J/\psi J/\psi $
reaction is studied within an effective Lagrangian approach and Breit-Wigner
formula. The numerical results show that the cross section of $X(6900)$ at
the c.m. energy of 6.9 GeV is much larger than that from the background
contribution. Moreover, we estimate dozens of signal events can be detected by D0 experiment,
which indicates that searching for the $X(6900)$ via
antiproton-proton scattering may be a very important and
promising way. Therefore, related experiments are suggested to be carried out.
\end{abstract}

\maketitle

\section{Introduction}\label{sec:intro}

In recent decades, more and more exotic hadron states have been observed
\cite{Tanabashi:2018oca}. These exotic hadron states are not only conducive to
the development of hadron spectrum, but also provides an important
opportunity for us to better understand the multiquark states and strong
interactions \cite{Liu:2019zoy,Guo:2019twa}. Although most of these exotic
states that have been observed are concentrated in the charm or bottom quark
energy region, the exotic states composed entirely of heavy quarks are still
very limited \cite{Tanabashi:2018oca,Liu:2019zoy,Guo:2019twa}.

Very recently, a narrow $X(6900)$ structure, which may consisting of four
charm quarks, was observed in $J/\psi $ pair invariant mass spectrum by LHCb
experiment with more than 5$\sigma $ of significance level \cite%
{Aaij:2020fnh}. The mass and width of the $X(6900)$\ resonance are measured
to be either%
\begin{eqnarray}
M &=&6886\pm 11\pm 11\text{ MeV}, \\
\Gamma &=&168\pm 33\pm 69\text{ MeV,}
\end{eqnarray}%
based on the simple model with interference. Actually, based on various
theoretical models, a lot of research has been conducted on fully-heavy
tetraquark states \cite%
{Chao:1980dv,Iwasaki:1975pv,Ader:1981db,Heller:1985cb,Badalian:1985es,Zouzou:1986qh,Lloyd:2003yc,Barnea:2006sd,Karliner:2016zzc,Chen:2016jxd,Bai:2016int,Wang:2017jtz,Debastiani:2017msn, Anwar:2017toa,Esposito:2018cwh,Wu:2016vtq,Hughes:2017xie,Richard:2018yrm,Wang:2019rdo,Chen:2019dvd,Liu:2019zuc,Deng:2020iqw,liu:2020eha,Lu:2020cns,Wang:2020ols,Chen:2020xwe,Yang:2020rih,Jin:2020jfc}%
, which are important to reveal the structure and properties of the
fully-heavy tetraquark state. For example, the $X(6900)$ was interpreted as
a $P-$wave tetraquark state in a nonrelativistic quark model \cite%
{liu:2020eha}, or the first radial excitation of $cc\bar{c}\bar{c}$ in an
extended relativized quark model \cite{Lu:2020cns}. In the QCD sum rule
framework, the narrow structure $X(6900)$ can be interpreted as a second
radial excited $S-$wave tetraquark state \cite{Wang:2020ols} or a $P-$wave
state \cite{Chen:2020xwe} with $J^{PC}=0^{-+},1^{-+}$, respectively. In
refs. \cite{Yang:2020rih,Jin:2020jfc}, the theoretical results indicate that
there may exist the resonance states, with masses range between 6.3 GeV to
7.4 GeV, and the quantum numbers $J^{P}=0^{+},1^{+}$ and $2^{+}$.
In ref. \cite{Mikhasenko:2020qor}, the spin parity of the state formed by the two-vector system was discussed,
and a possible method for determining its quantum number was given.

In addition to the study on the structure and properties of $X(6900)$, the
research on the production of the $X(6900)$ through different scattering
reactions is also very important and will help us to determine its nature as
genuine states. For example, in ref. \cite{Becchi:2020uvq}, the production
of the ground $cc\bar{c}\bar{c}$ with $J^{PC}=0^{++},2^{++}$ in $pp$
collisions was calculated, and the upper limit of cross section is about 40
fb for the 4 muons channel. In addition to $pp$ collisions, determining the
tetraquark state via the annihilation of positive and negative protons is
usually an important and effective way \cite%
{Chen:2008cg,Lange:2013sxa,Zotti:2014hza,Lin:2012ru,Wang:2015pfa,Wang:2015uua,Abazov:2015sxa}%
. In this work, the discovery potential for the tetraquark state $X(6900)$
via $\bar{p}p$ annihilation reaction will be investigated. One will estimate
the cross-section of $X(6900)$ production via $\bar{p}p\rightarrow J/\psi
J/\psi $ reaction and analyze the corresponding background contribution. The
theoretical results obtained will be an important theoretical basis for the $%
\bar{p}p$ annihilation experiment.

This paper is organized as follows. After the introduction, one present the
formalism for the production of $X(6900)$ in Section II. The numerical
results of the $X(6900)$ production follow in Section III. Finally, the paper%
{\ ends} with a brief summary.

\section{Formalism}

\subsection{$X(6900)$ production in $\bar{p}p$ annihilation}

\begin{figure}[tbph]
\begin{center}
\includegraphics[scale=0.5]{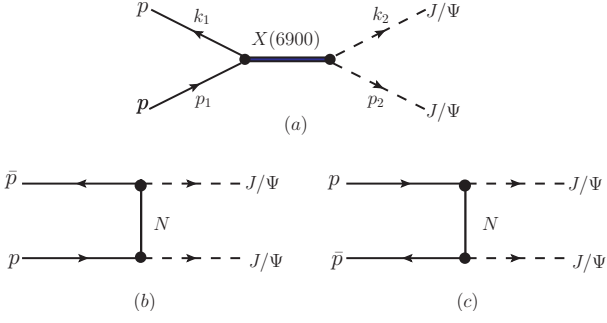}
\end{center}
\caption{Feynman diagrams for the reaction $\bar{p}p\rightarrow
X(6900)\rightarrow \protect\psi \protect\psi $.}
\label{fig1}
\end{figure}

The Feynman diagram of the $\bar{p}p\rightarrow J/\psi J/\psi $ reaction via
$s$ channel $X(6900)$ exchange is depicted in Fig. \ref{fig1} (a). For
convenience, $X(6900)$ and $J/\psi $ will be abbreviated as $X$ and $\psi $,
receptively.

\

For the production of the resonance $X(6900)$ in $s$-channel of $\bar{p}%
p\rightarrow \psi \psi $ reaction, the cross section can be calculated by
the standard Breit-Wigner formula \cite{Tanabashi:2018oca}.%
\begin{equation}\label{eq:csx}
\mathcal{\sigma }_{X}=\frac{2J+1}{(2S_{1}+1)(2S_{2}+1)}\frac{4\pi }{%
k_{in}^{2}}\frac{\Gamma ^{2}}{4}[\frac{Br(X\rightarrow \bar{p}%
p)Br(X\rightarrow \psi \psi )}{(W-W_{0})^{2}+\Gamma ^{2}/4}],
\end{equation}%
Where $W$ is the c.m. energy, $J$ is the spin of the resonance $X(6900)$,
and the $S_{1}$ and $S_{2}$ are the spin of initial anti-proton and proton,
respectively. Moreover, $k_{in}$ is the c.m. momentum in the initial state, $%
W_{0}$ is the c.m. energy at mass of $X(6900)$, and $\Gamma $ is the full
width of $X(6900)$.

The production cross-section of the $X(6900)$ structure relative to that of
all $J/\psi $ pair, times the branching fraction $Br(X\rightarrow \psi \psi
) $, $\mathcal{R}$, is determined as $[2.6\pm 0.6$ (stat) $\pm $ 0.8 (syst)$%
] $ by LHCb experiment \cite{Aaij:2020fnh}. Therefore, in this work, we
roughly take the value of the branching ratio $Br(X\rightarrow \psi \psi )$ $%
\simeq 2$.

Having fixed the branch ratio $Br(X\rightarrow \psi \psi )$, we turn to the value of $Br(X\rightarrow \bar{p}p)$, which is indispensable in Eq. \ref{eq:csx}. However, it has not been well determined, neither experimentally nor theoretically. Meanwhile, we notice in refs. \cite{Lange:2013sxa,Zotti:2014hza} that, the branching fraction of $%
Br(Y(4260)\rightarrow \bar{p}p)$ and $Br(\psi (4040)\rightarrow \bar{p}p)$
were estimated using the branching ratio of $J/\psi $ state, multiplying by
the ratio of the width of the $Y(4260)$ or $\psi (4040)$ to the width of $%
J/\psi $. Thus in this work, we will adopt the same
method as in refs. \cite{Lange:2013sxa,Zotti:2014hza} to naively determine the
branching ratio of $X$ decaying into $\bar{p}p$. Due to the lack of deep understanding of the full-charm tetraquark $X(6900)$, its inner structure and quantum numbers are still unknown. Different $J^{PC}$ assignments were assumed to evaluated the mass of a full-charm tetraquark, as discussed in Sec. \ref{sec:intro}. To carry out the estimation of the cross section of the $s$-channel, we assume
that the quantum number $J^{PC}$ of X(6900) is $0^{++}$ or $2^{++}$, same as the assumption in Ref. \cite{Becchi:2020uvq}. In the
known charmonium family, the spin-parity of $\chi _{c0}$ and $\chi _{c2}$ are $%
0^{++}$ and $2^{++}$, respectively. Therefore, we plan to simply replace $%
J/\psi $ with $\chi _{c0}$ and $\chi _{c2}$ to estimate the branching ratio
of $X\rightarrow \bar{p}p$, namely,%
\begin{equation}
Br(X\rightarrow \bar{p}p)\simeq Br(\chi _{c0}\rightarrow \bar{p}p)\times \frac{%
\Gamma _{\chi _{c0}}}{\Gamma _{X}},\text{ for }X\text{ with }J^{PC}=0^{++}
\end{equation}
and%
\begin{equation}
Br(X\rightarrow \bar{p}p)\simeq Br(\chi _{c2}\rightarrow \bar{p}p)\times \frac{%
\Gamma _{\chi _{c2}}}{\Gamma _{X}},\text{ for }X\text{ with }J^{PC}=2^{++}
\end{equation}%
where $\Gamma _{\chi _{c0}}=10.8$ MeV, $\Gamma _{\chi _{c2}}=1.97$ MeV and $%
\Gamma _{X}=168$ MeV are the total width of $\chi _{c0}$, $\chi _{c2}$ and $%
X(6900)$ state, respectively. By taking the branching ratio $Br(\chi
_{c0}\rightarrow \bar{p}p)=2.24\times 10^{-4}$ and $Br(\chi _{c2}\rightarrow
\bar{p}p)=7.33\times 10^{-5}$, one get $Br(X\rightarrow \bar{p}p)=1.44\times
10^{-5}$ for $X$ with $J^{PC}=0^{++}$ and $Br(X\rightarrow \bar{p}%
p)=8.6\times 10^{-7}$ for $X$ with $J^{PC}=2^{++}$. Here, it must be noted
that if we use the branching ratio of $J/\psi $ for estimation, $%
Br(X\rightarrow \bar{p}p)=1.17\times 10^{-6}$ is obtained, which is just
between the value estimated by the branching ratio of $\chi _{c0}$ and $\chi
_{c2}$.

\subsection{The background analysis}

Fig. \ref{fig1} (b)-(c) presents the $\bar{p}p\rightarrow \psi \psi $
reaction with $t$ and $u$ channel by exchanging a nucleon, which can be
considered as the main background contributions for the production of $%
X(6900)$. By employing the effective Lagrangian approach, the cross section
of $\bar{p}p\rightarrow \psi \psi $ reaction can be calculated.

The Lagrangian density for the vertice of $\psi NN$ is written as \cite%
{Lin:2012ru},

\begin{equation}
\mathcal{L}_{\psi NN}=-g_{_{\psi NN}}\bar{N}\gamma ^{\mu }N\psi _{\mu },
\end{equation}%
where $\psi $ and $N$ denote the fields of $J/\psi $ and nucleon,
respectively.

The values of coupling constant $g_{_{\psi NN}}$ can be derived from the
corresponding decay width,

\begin{equation*}
\Gamma _{\psi \rightarrow \bar{p}p}=\left( g_{\psi NN}\right) ^{2}\frac{%
\sqrt{m_{\psi }^{2}-4m_{N}^{2}}}{12\pi m_{\psi }^{2}}\left( m_{\psi
}^{2}+2m_{N}^{2}\right)
\end{equation*}%
Thus we get $g_{_{\psi NN}}\simeq 1.6\times 10^{-3}$, which is calculated by
the measured branching fractions and total widths of $J/\psi $ ($m_{\psi
}=3096.916$ MeV and $\Gamma _{\psi \rightarrow \bar{p}p}\simeq 0.197$ keV)
\cite{Tanabashi:2018oca}.

Based on the Lagrangians above, the scattering amplitude for the reactions $%
\bar{p}p\rightarrow \psi \psi $ can be constructed as%
\begin{equation}
-i\mathcal{M}_{\bar{p}p\rightarrow \psi \psi }=\epsilon _{\psi }^{\nu
}(p_{2})\bar{u}(p_{1})\mathcal{A}_{\mu \nu }u(k_{1})\epsilon _{\psi }^{\mu
}(k_{2}),
\end{equation}%
where $u$ is the Dirac spinor of nucleon, and $\epsilon _{\gamma }$ is the
polarization vector of photon.

The reduced amplitude $\mathcal{A}_{\mu \nu }$ for the $t$ and $u$ channel
background reads
\begin{eqnarray}
\mathcal{A}_{\mu \nu }^{t} &=&(g_{\psi NN}g_{\psi NN})\gamma _{\nu }\frac{%
\rlap{$\slash$}q_{t}+m_{N}}{q_{t}^{2}-m_{N}^{2}}\gamma _{\mu }\mathcal{F}%
_{t}^{2}\mathcal{(}q_{t}^{2}\mathcal{)}, \\
\mathcal{A}_{\mu \nu }^{u} &=&(g_{\psi NN}g_{\psi NN})\gamma _{\mu }\frac{%
\rlap{$\slash$}q_{u}+m_{N}}{q_{u}^{2}-m_{N}^{2}}\gamma _{\nu }\mathcal{F}%
_{u}^{2}\mathcal{(}q_{u}^{2}\mathcal{)},
\end{eqnarray}

For two $\psi NN$ vertices, a general form factor $\mathcal{F}%
_{t}(q_{t}^{2})=\mathcal{F}_{u}\mathcal{(}q_{u}^{2}\mathcal{)=}(\Lambda
_{t/u}^{2}-m_{N}^{2})/(\Lambda _{t/u}^{2}-q_{N}^{2})$ is taken into account
\cite{Lin:2012ru,Wang:2015uua}. In refs. \cite{Lin:2012ru,Wang:2015uua}, it
can be found that the cross section of $\bar{p}p\rightarrow J/\psi \pi $
reaction with nucleon exchange were consistent with the E760 and E835 data
by taking $\Lambda _{t/u}=1.9$ and $3.0$ GeV, respectively. In the spirit of
estimating the upper limit of background contribution, in this work, we take
$\Lambda _{t}=\Lambda _{u}=3$ GeV.

With the preparation in the previous section, the cross section of the
reaction $\bar{p}p\rightarrow \psi \psi $ can be calculated. The
differential cross section in the center of mass (c.m.) frame is written as
\begin{equation}
\frac{d\sigma }{d\cos \theta }=\frac{1}{32\pi s}\frac{\left\vert \vec{k}%
_{2}^{{~\mathrm{c.m.}}}\right\vert }{\left\vert \vec{k}_{1}^{{~\mathrm{c.m.}}%
}\right\vert }\left( \frac{1}{4}\sum\limits_{\lambda }\left\vert \mathcal{M}%
\right\vert ^{2}\right) ,
\end{equation}%
Here, $s=(k_{1}+p_{1})^{2}$, and $\theta $ denotes the angle of the outgoing
$J/\psi $ meson relative to $\bar{p}$ beam direction in the c.m. frame. $%
\vec{k}_{1}^{{~\mathrm{c.m.}}}$ and $\vec{k}_{2}^{{~\mathrm{c.m.}}}$ are the
three-momenta of the initial photon beam and final $J/\psi $ meson,
respectively.

\section{Numerical results}

\begin{figure}[h]
\begin{center}
\includegraphics[scale=0.41]{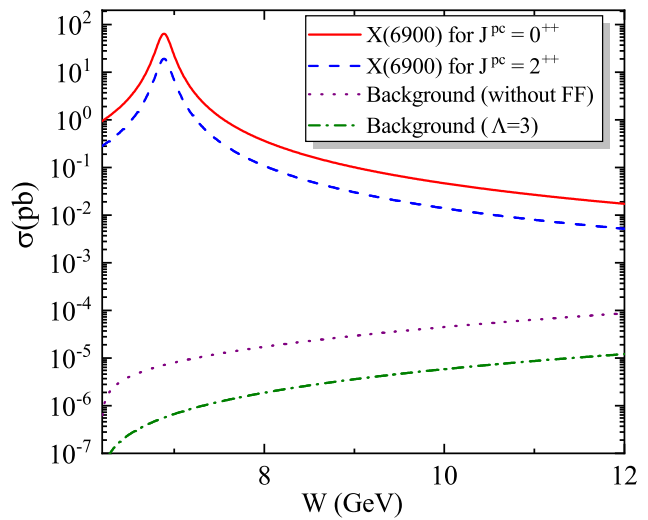}
\end{center}
\caption{The total cross section for the $\bar{p}p\rightarrow \psi \psi $
reaction. The red solid and blue
dashed lines are for the cross section of $X(6900)$ production when the spin-parity is 0++ and 2++,
respectively. Moreover, the green dot-dashed and violet dotted are for the contributions from background with and without
form factor, respectively.}
\label{tcs}
\end{figure}

After the above preparations, one calculated the total cross section for the
reaction $\bar{p}p\rightarrow \psi \psi $ from threshold to 12 GeV of the
center of mass energy, as depicted in Fig. \ref{tcs}. As can be seen from
the Fig. \ref{tcs}, the cross section from the $X(6900)$ contribution have a
distinct peak near the center of mass energy of 6.9 GeV. And when the
spin-parity quantum number of $X$ is $0^{++}$ or $2^{++}$, the cross section
of $X$ production through the $s$ channel can reach dozens of pb, which is
much larger than the cross section from the background contribution. In
addition, we also calculated the background cross section without the from
factor (abbreviated as FF). Although it is more than an order of magnitude
higher than the background cross section with the FF added, it is still much
lower than the cross section of the $X$ production. The main reason for the
very small background term is that the $g_{_{\psi NN}}$ coupling constant is
very small.

The above results indicate that the best energy window for searching for the
fully-charm tetraquark state $X(6900)$ via the $\bar{p}p\rightarrow \psi
\psi $ process is near the c.m. energy 6.9 GeV, in which the signal can be
clearly distinguished from background. The D0 \cite{Abazov:2015sxa} and the forthcoming PANDA\cite{Lange:2013sxa} experiments are ideal platforms to study new hadronic states via $\bar{p}p$ reactions, of which the designed c.m. energy is below the double $J/\psi$ threshold for the latter one. By taking the cross section of $X(6900)$ production calculated above, one finds that the number of events of $X(6900)$ can reach as many as dozens if taking an integrated luminosity of 10.4 fb$^{-1}$ collected with the D0 detector, when a 50\% selection efficiency is adopted. Considering that the background is very clean, it should be possible for $X(6900)$ to be observed by the D0 experiment with a high confidence level. Moreover, in this work the cross section of $
X(6900)$ is obtained when the quantum number is assumed to be $0^{++}$ or $2^{++}$. In fact, the quantum number of $X(6900)$ is currently undecided.
But according to our estimation, even if $X(6900)$ takes other quantum
numbers, its cross section is about the same order as the cross section we
have obtained so far. This seems to mean that no matter what the quantum
number of $X$ is, it may be an effective way to find or observe $X(6900)$
through the $\bar{p}p$ annihilation.

\section{Summary and discussion}

In this work, the $X(6900)$ production in the $\bar{p}p\rightarrow \psi \psi
$ reaction is investigated by employing the effective field theory and the
Breit-Wigner formula. The numerical results show that the cross section of $X
$ production via the $s$ channel can reach dozens of pb at the best energy
window $W=6.9$ GeV, which the signal can be clearly distinguished from
background. In addition, according to our estimation, dozens of $X(6900)$ can be detected with a data sample of 10.4 fb$^{-1}$ collected with the D0 detector, which indicates that it is
feasible to find $X(6900)$ through the $\bar{p}p\rightarrow \psi \psi $
reaction. Hence, an experimental study of fully charm tetraquark $X(6900)$
via $\bar{p}p$ annihilation is suggested, which will be of great
significance to clarify the production mechanism and nature of $X(6900)$.

It should be noted that the $p\bar{p}$ will be annihilated into gluons first, then those gluons will be converted into the di-$J/\psi$ final state. The production of the full-charm tetraquarks at LHC was discussed and the relative cross section for the process $gg\to J/\psi J/\psi$ was estimated \cite{Berezhnoy:2011xy,Berezhnoy:2011xn}. The cross sections performed are around 10 pb, which have the same order of magnitude with the results exhibited here.

\section{Acknowledgments}

This project is supported by the National Natural Science Foundation of
China (Grant Nos. 11705076 and 11747160), and by the Strategic Priority
Research Program of Chinese Academy of Sciences, Grant No. XDB34030301. This
work is partly supported by HongLiu Support Funds for Excellent Youth
Talents of Lanzhou University of Technology. We also acknowledgment
the Fundamental Research Funds for the Central Universities (Grant
Nos. 31020180QD118 and 2682020CX70), the Development and Exchange Platform
for Theoretic Physics of Southwest Jiaotong University in 2020 (Grant No. 11947404).

\end{document}